\documentclass[final,5p,times,twocolumn,sort&compress]{elsarticle}
\usepackage{lineno,hyperref}
\usepackage{graphicx,graphics}
\usepackage{amsmath}
\usepackage{amssymb}
\usepackage{epstopdf}
\usepackage{amstext}
\usepackage{amsthm}
\usepackage{amsfonts}
\usepackage{latexsym}
\usepackage{array}
\usepackage{xfrac}
\usepackage{color}
\usepackage{subfigure}
\usepackage{multirow}
\usepackage{fontenc}
\usepackage{bm}
\usepackage{adjustbox}
\usepackage{dcolumn}
\usepackage{color}
\modulolinenumbers[5]
\begin{document}
\begin{frontmatter}

\title{New Gogny interaction suitable for astrophysical applications}

\author[BCN]{C. Gonzalez-Boquera}
\author[BCN]{M. Centelles}
\author[BCN]{X. Vi\~nas}
\author[Autonoma,Poli]{L.M. Robledo}

\address[BCN]{Departament de F\'isica Qu\`antica i Astrof\'isica and Institut de Ci\`encies del Cosmos (ICCUB), 
Facultat de F\'isica, Universitat de Barcelona, Mart\'i i Franqu\`es 1, E-08028 Barcelona, Spain}
\address[Autonoma]{Departamento de F\'isica Te\'orica,
Facultad de F\'isica, Universidad Aut\'onoma de Madrid,E-28049 Madrid, Spain} 
\address[Poli]{Center for Computational Simulation,
Universidad Polit\'ecnica de Madrid,
Campus de Montegancedo, Boadilla del Monte, 28660-Madrid.}

\date{\today}

\begin{abstract}
The D1 family of parametrizations of the Gogny interaction commonly suffers from
a rather soft neutron matter equation of state that leads 
to maximal masses of neutron stars well below the observational value of two solar masses.
We propose a reparametrization scheme that preserves the good 
properties of the Gogny force but allows one to tune the density dependence 
of the symmetry energy, which, in turn, modifies the predictions for the maximum stellar mass. 
The scheme works well for D1M, and leads to a new parameter set, dubbed D1M$^{*}$. In 
the neutron-star domain, D1M$^{*}$ predicts a maximal mass of two solar masses
and global properties of the star in harmony with those obtained with the 
SLy4 Skyrme interaction. By means of a set of selected calculations in finite nuclei, 
we check that D1M$^{*}$ performs comparably well to D1M in several aspects of nuclear 
structure in nuclei.
\end{abstract}





\end{frontmatter}

\section{Introduction}
Neutron stars (NSs) are among the densest objects in the Universe. 
From the surface to the center of a NS the density varies by about
fifteen orders of magnitude, involving several physical scenarios \cite{Shapiro83,Haensel06}.
The outermost part of the star, or outer crust, consists of ionized atomic nuclei 
embedded in a free electron gas. These nuclei arrange themselves in a solid 
lattice to minimize the Coulomb repulsion and are stabilized against $\beta$-decay by the electron gas \cite{Haensel06,Baym71}.
In the deepest layers of the outer crust, nuclei become so neutron rich that neutrons 
start to drip.
Hence, the structure of the inner crust 
consists of a Coulomb lattice of nuclear clusters permeated by free neutron and electron gases  
(see e.g.\ \cite{Negele73,Douchin01,Sharma15} and refs.\ therein). At the bottom 
of this region, the nuclear clusters may adopt non-spherical shapes (``nuclear pasta'')
in order to minimize the Coulomb energy. The inner crust extends up to densities about one half of the 
saturation density of nuclear matter ($\simeq$1.3$\times 10^{14}$ g/cm$^3$). 
In the interior region of a NS, or core, matter forms a homogeneous liquid composed of neutrons 
plus a certain fraction of protons, electrons and muons, and eventually other exotic particles,
under $\beta$-equilibrium and charge neutrality \cite{Shapiro83,Haensel06}.
This region accommodates most of the mass and size of the star, implying that 
global properties such as the maximum mass, radius, or moment of inertia of the NS 
are determined to a large extent by the properties of the homogeneous core.

From a theoretical point of view, the essential ingredient to study NSs is the equation of state (EOS) of matter \cite{Oertel:2016bki}.
Although nowadays sophisticated ab initio calculations can be performed to describe the homogeneous matter in the core,
effective nuclear interactions such as Skyrme forces or the relativistic mean field (RMF) theory, 
which successfully describe many properties of terrestrial nuclei, are in wide use for NS
calculations due to their relative simplicity.
For instance, Skyrme-HFB models \cite{Goriely10} have reached a high degree of accuracy in predicting 
experimental masses and are at the same time well suited for astrophysical studies \cite{Chamel11,Fantina13}. 
More occasionally, also Gogny forces have been applied in NS calculations \cite{Than11,Doan11,Rios14,Gonzalez17}. 
Due to the complexity of modeling the inner crust, there are few EOSs that describe 
the whole NS from the crust to the core in a unified manner. 
Some examples are, among others \cite{Oertel:2016bki}, 
the EOS of Lattimer-Swesty \cite{Lattimer91}, the SLy EOS of Douchin-Haensel \cite{Douchin01},
the EOSs of the BSk family developed by the Brussels group \cite{Chamel11,Fantina13},
which are based on Skyrme forces, the EOS by Shen et al.\ \cite{Shen98} obtained within the RMF theory,
or the BCPM EOS based on Brueckner calculations \cite{Sharma15}.

In Skyrme forces the interaction is of zero range and the pairing force, which is needed to 
study open-shell nuclei, is not connected with the force used to describe the mean field. 
The Gogny interaction was 
proposed more than thirty years ago aimed to describe the mean field and the pairing field simultaneously with the same 
finite-range effective force \cite{Gogny80}. 
Large-scale HFB calculations based on the D1S Gogny parametrization \cite{Berger91} revealed  
some deficiencies in the description of nuclear masses compared to experimental data. 
To overcome these limitations, new Gogny parametrizations such as D1N \cite{Chappert08} 
and D1M \cite{Goriely09} have been proposed. At variance with D1S and D1N, which follow the D1 fitting protocol \cite{Gogny80}, 
the D1M force has been fitted by minimizing the difference to 2149 measured nuclear masses \cite{Audi12} 
and including quadrupole correlation energies.
 It is important to mention that in the calibration of D1N and D1M, 
the energy of neutron matter is required to qualitatively reproduce the microscopic 
calculations of Friedman and Pandharipande \cite{Friedman81}. D1M reproduces  
the 2149 experimental masses with a rms deviation as low as 798 keV \cite{Goriely09}. 
Unfortunately, the extrapolation to the domain of neutron stars with the Gogny parametrizations
works less well. It has been found \cite{Doan11,Rios14,Gonzalez17} that the most successful 
Gogny forces for describing finite nuclei, namely D1S, D1N and D1M, are unable to reach NS masses of about 
$2 M_\odot$, as required by recent astrophysical observations \cite{Demorest10,Antoniadis13}. Moreover, only
few Gogny forces, including D1M but not D1N nor D1S, achieve a NS mass above the canonical 
$1.4 M_\odot$ value \cite{Rios14,Gonzalez17}. Therefore, the aim of this work is to introduce a new Gogny 
force, which we call D1M$^*$, that retaining a similar quality to D1M for finite nuclei, 
may be used to study NS physics at a level of the most successful Skyrme forces.
We next analyze the neutron-star matter EOS and the predictions for NS masses and radii provided by different Gogny forces, 
and, specially, by D1M$^*$. The fit of the new force D1M$^*$ is discussed afterwards, and its ability for describing infinite 
nuclear matter and finite nuclei is investigated.

\section{Neutron-star matter described with Gogny interactions}
\label{sec:star}
The standard Gogny interaction of the D1 family consists of a finite-range part, which is modeled by two
Gaussian terms including all the possible spin-isospin exchange terms, plus a 
\mbox{zero-range density-dependent term. Adding the spin-orbit force,} 
which is also of contact type, the Gogny interaction reads \cite{Gogny80}:
\begin{eqnarray}\label{VGogny}
  V (\mathbf{r}_1 , \mathbf{r}_2) &=& \sum_{i=1,2} 
  \big( W_i + B_i P_{\sigma} - H_i P_{\tau} - M_i P_{\sigma}P_{\tau}\big)
   e^{-r^2/\mu_i^{2}} \nonumber
   \\
  && \mbox{} + t_3 (1+ x_3 P^\sigma) \rho^\alpha (\mathbf{R}) \delta (\mathbf{r}) \nonumber
 \\
  && \mbox{} + i W_{LS} (\sigma_1 + \sigma_2) (\mathbf{k'} \times \delta (\mathbf{r}) \mathbf{k}), 
\end{eqnarray}
where ${\bf r}$ and ${\bf R}$ are the relative and the center of mass coordinates of the two nucleons, and 
$\mu_1 \simeq 0.5$--0.7 fm and $\mu_2 \simeq 1.2$ fm are the ranges of the two Gaussian form
factors, which simulate the short- and long-range components of the force, respectively. In 
Eq.~(\ref{VGogny}), $\mathbf{k}= (\overrightarrow{\nabla}_1-\overrightarrow{\nabla}_2)/2i$ is the
relative momentum between the two nucleons and $\mathbf{k}'$ is its complex conjugate.

The symmetry energy is the basic quantity that rules the isovector part of the interaction. 
It is obtained as $E_{\rm sym} (\rho) = \frac{1}{2} \partial^2 E_b (\rho, \delta)/\partial \delta^2 \vert_{\delta=0}$ from the 
energy per particle $E_b (\rho, \delta)$ in asymmetric nuclear matter of density $\rho = \rho_n + \rho_p$ and isospin asymmetry
$\delta = (\rho_n - \rho_p)/\rho$, where $\rho_n$ and $\rho_p$ are the neutron and proton densities. In the Gogny interaction,
the symmetry energy becomes~\cite{Gonzalez17}:
\begin{eqnarray}
&&\hspace*{-12mm} E_{\rm sym}(\rho) = \frac{\hbar^2}{6m} \left( \frac{3 \pi^2}{2}\right)^{2/3} \rho^{2/3} - \frac{t_3}{8} \rho^{\alpha + 1}
(2 x_3 +1) - \frac{1}{6\sqrt{\pi}} \times  \nonumber \\
&&\hspace*{-12mm} \sum_{i=1,2} \Bigg\{ \mu_i^3 k_F^3 \left( H_i + \frac{M_i}{2} \right) + \frac{\left( 1 - e^{-\mu_i^2 k_F^2} \right)}{\mu_i k_F} 
\left( W_i + 2 B_i - 2 H_i -4 M_i\right)  \nonumber \\
&&\hspace*{-6mm} \mbox{} + \mu_i k_F \left[ H_i + 2 M_i -  e^{-\mu_i^2 k_F^2} \left( W_i + 2 B_i - H_i - 2 M_i \right)\right] \Bigg\} ,
\label{eq-esym}
\end{eqnarray}
where $k_F=(3\pi^2\rho/2)^{1/3}$ is the Fermi momentum.
The slope parameter $L$, defined as $L=3\rho_0 \partial E_{\rm sym}(\rho)/\partial \rho \vert_{\rho=\rho_0}$ where $\rho_0$ is the 
saturation density, provides a good handle on the density dependence of the symmetry energy around saturation. The $L$ value is known to be 
strongly correlated with the isospin properties, such as e.g.\ neutron densities and neutron skins, of nuclei \cite{Li:2014oda}.

\begin{figure}[t]
\centering
\includegraphics[width=0.9\columnwidth,clip=true]{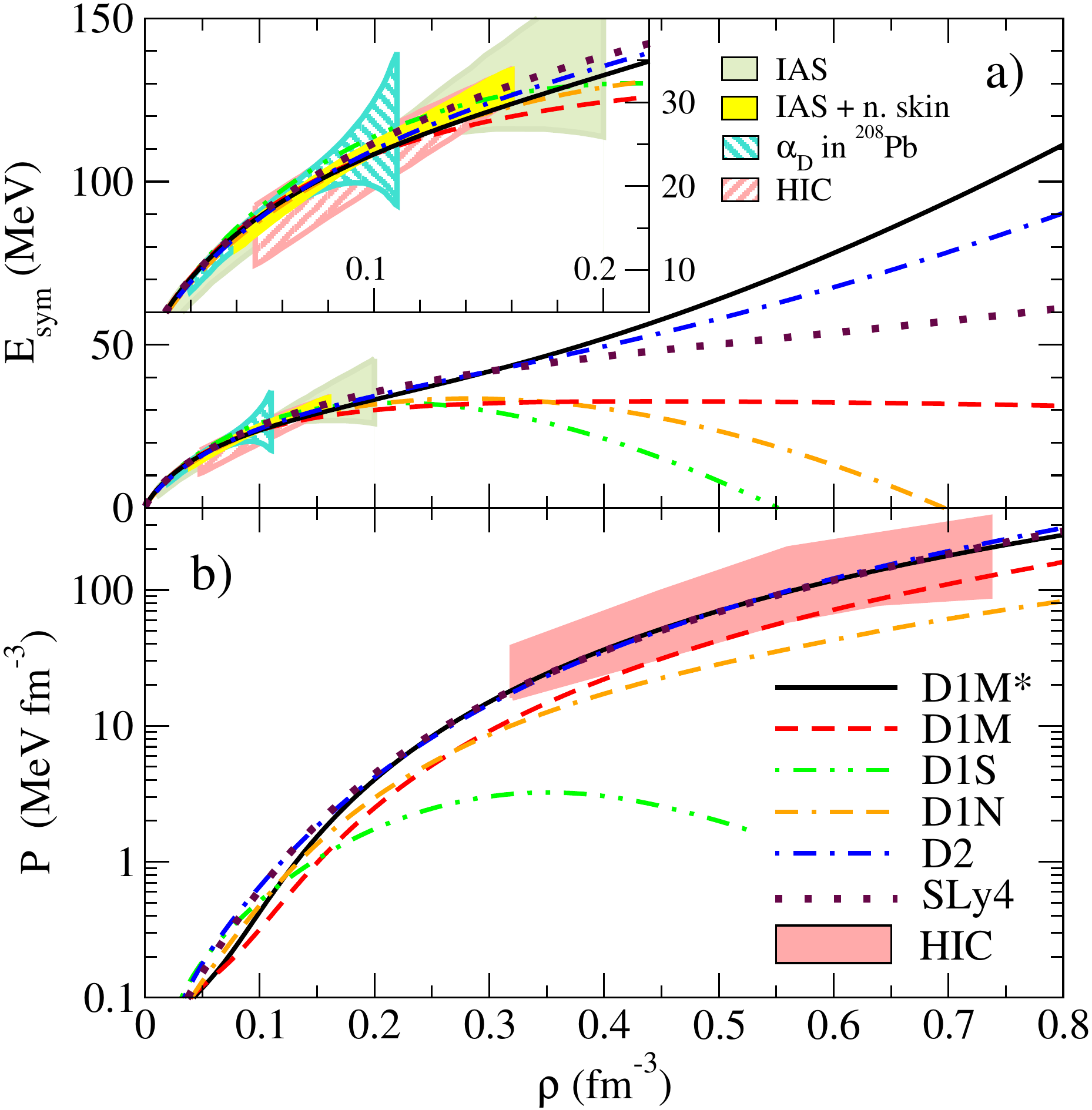}
\caption{a) Symmetry energy versus density from the D1S, D1N, D1M, D1M$^*$ and D2 Gogny forces 
and from the SLy4 Skyrme force. The inset is a magnified view of the low-density region.
Also plotted are the constraints from isobaric analog states (IAS) and from IAS and neutron skins (IAS+n.skin) \cite{Danielewicz14}, 
from the electric dipole polarizability in lead ($\alpha_D$ in $^{208}$Pb) \cite{Zhang:2015ava}
and from transport in heavy-ion collisions (HIC) \cite{Tsang:2008fd}.
b) Pressure in $\beta$-stable nuclear matter in logarithmic scale
as a function of density for the same interactions of panel~a). The shaded area depicts the
region compatible with collective flow in HICs~\cite{Danielewicz:2002pu}.\label{esym}}
\end{figure}

The symmetry energy as a function of density is displayed in Fig.~\ref{esym}a for several
Gogny forces and for the SLy4 Skyrme force, used in the NS SLy EOS of Douchin-Haensel \cite{Douchin01}. 
At subsaturation densities the symmetry energy in the considered forces displays a similar behavior  
and takes a value of about 30 MeV at saturation. 
The subsaturation regime is also the finite nuclei regime, where the parameters of the nuclear forces are fitted to.
Indeed, we observe in Fig.~\ref{esym}a that at subsaturation the present forces fall within or are very close to the region 
compatible with recent constraints on $E_{\rm sym} (\rho)$ deduced from several nuclear observables \cite{Danielewicz14,Zhang:2015ava,Tsang:2008fd}.
Above saturation, in contrast, the behavior of the calculated symmetry energy shows a strong model dependence. 
The Gogny parametrizations usually extrapolate to high density with a too soft symmetry energy.
For example, D1M shows a nearly flat behavior at suprasaturation, and the $E_{\rm sym} (\rho)$ curves of D1S and D1N, after 
reaching a maximum at $\rho\sim 0.2$--0.3 fm$^{-3}$, bend down until they become negative at some density a few times the saturation one, 
which indicates the onset of an isospin instability. 
Although this happens at large densities for terrestrial phenomena, it is critical for neutron stars, 
where larger densities occur in the star's interior. 
The other Gogny forces in the figure, i.e., D2 \cite{Chappert15} and the new D1M$^*$ force of this work,
exhibit an increasing $E_{\rm sym} (\rho)$ with growing density and do not present the isospin instability.
D2 is a very recent Gogny interaction \cite{Chappert15} devised by the Bruy\`eres-le-Ch\^atel 
group, where the usual zero-range density-dependent term of the D1 family is replaced by a finite-range term.
As in the D1N and D1M cases, the fit of D2 requires the 
reproduction of the microscopic energy of neutron matter \cite{Friedman81}.
Concerning the results for finite nuclei, D2 \cite{Chappert15} describes the binding energies along isotopic chains 
without the drift of the energies with increasing neutron number observed in D1S \cite{Chappert07}. 
However, as pointed out in \cite{Chappert15}, the global description of nuclear masses using the D2 force 
does not reach yet the quality obtained with D1M \cite{Goriely09}.
The D1M$^*$ force is a new Gogny parametrization of the type of Eq.~(\ref{VGogny}), introduced here for the first time. 
It is devised to keep the quality of the description of finite nuclei at the level 
of D1M and to be able to predict NSs fulfilling the astrophysical observations of two solar masses. Details on the strategy 
followed to obtain D1M$^*$ and the parameters of this force are given later in Sec.~\ref{sec-fit}.

The mass-radius (M-R) relation in neutron stars is dictated by the corresponding EOS, which is the essential
ingredient to solve the Tolman-Oppenheimer-Volkov (TOV) equations \cite{Shapiro83}. 
The EOS (total pressure against density) of $\beta$-stable, globally charge-neutral NS matter \cite{Rios14,Gonzalez17}
calculated with the given functionals is displayed in Fig.~\ref{esym}b. 
The new Gogny force D1M* and D2 predict a high-density EOS with a similar stiffness to 
the SLy4 EOS and they agree well with the region constrained by collective flow in energetic 
heavy-ion collisions (HIC) \cite{Danielewicz:2002pu}, shown as the shaded area in 
Fig.~\ref{esym}b.\footnote{Though the constraint of \cite{Danielewicz:2002pu} was proposed for neutron matter,
at these densities the pressures of $\beta$-stable matter and neutron matter are very similar.}
The EOSs from the original D1M parametrization and from D1N are significantly softer.
The D1S force yields a too soft EOS soon after saturation density, which implies it is not 
suitable for describing NSs.

\begin{figure}[t!]{}
\centering
\includegraphics[width=0.9\columnwidth,clip=true]{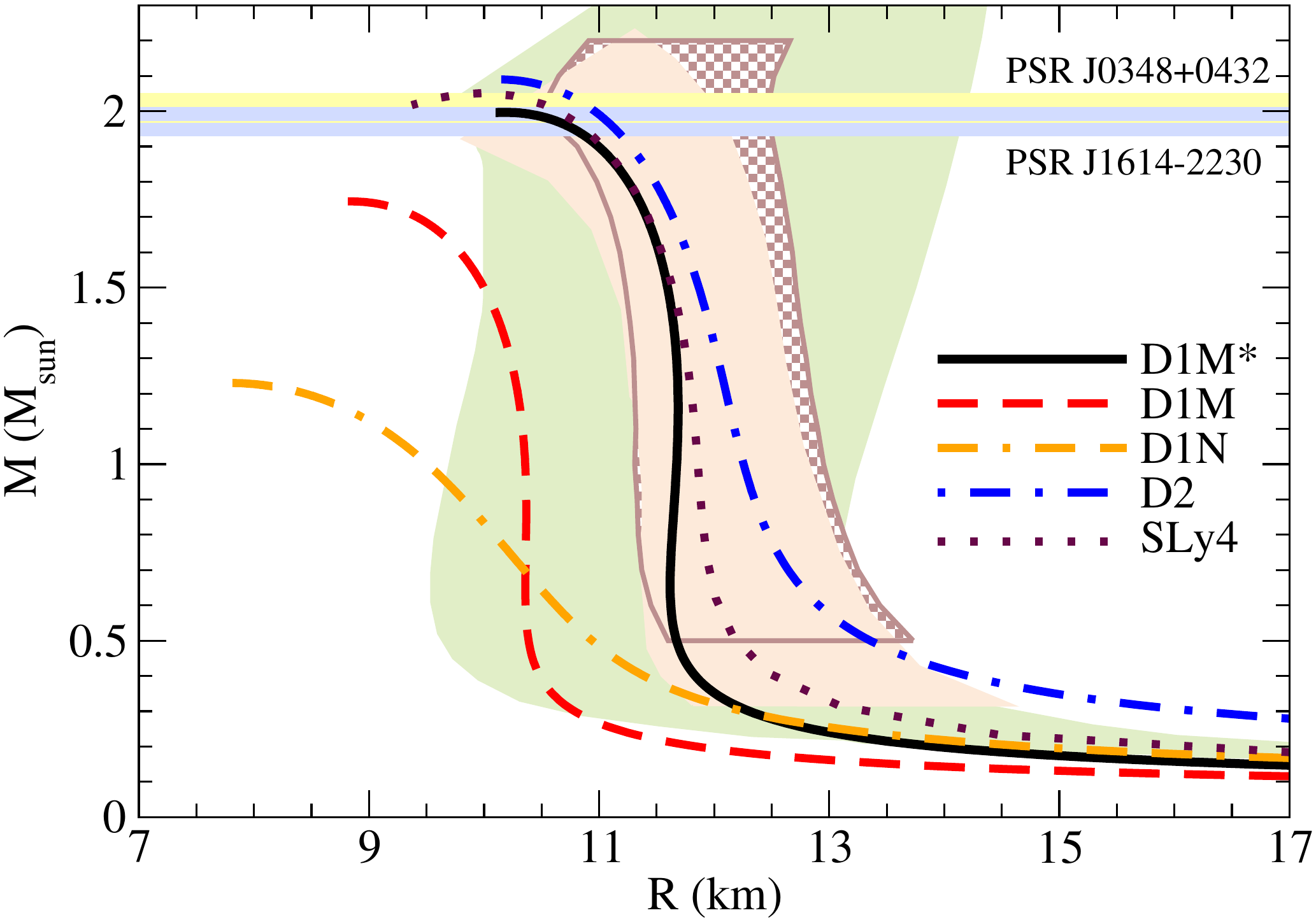}
\caption{Mass-radius relation in neutron stars from the D1N, D1M, D1M$^*$, D2 Gogny forces and the SLy4 Skyrme 
force.~The horizontal bands depict the heaviest observed NS masses \cite{Demorest10, Antoniadis13}.~The 
vertical green band shows the \mbox{M-R} region deduced from chiral nuclear interactions up to normal density 
plus astrophysically constrained high-density EOS extrapolations \cite{Hebeler13}. 
The brown dotted band is the zone constrained by the cooling tails of type-I X-ray bursts 
in three low-mass X-ray binaries and a Bayesian analysis \cite{Nattila16}, and the beige 
constraint at the front is from five quiescent low-mass X-ray binaries and five photospheric
radius expansion X-ray bursters after a Bayesian analysis~\cite{Lattimer14}.}
 \label{MR}
\end{figure}

To solve the TOV equations for a NS, knowledge of the EOS from the center to the surface 
of the star is needed. 
At present we do not have microscopic calculations of the EOS of the inner crust with 
Gogny forces.
In this work, following previous literature \cite{Link1999,Carriere03,Xu09,Gonzalez17}, we interpolate 
the inner-crust EOS by a polytropic form  $P = a + b \epsilon^{4/3}$ ($\epsilon$ is the 
mass-energy density), where the index $4/3$ assumes that the pressure at these densities 
is dominated by the relativistic degenerate electrons. 
We match this formula continuously to our Gogny EOSs of the homogeneous core
and to the Haensel-Pichon EOS of the outer crust \cite{Douchin01}. 
The core-crust transition density is selfconsistently computed for each Gogny force by the thermodynamical method \cite{Gonzalez17}. 
We show in Fig.~\ref{MR} the obtained NS mass-radius plots (results for D1S are not shown 
because D1S did not produce stable solutions of the TOV equations).
We also plot as a benchmark the M-R curve calculated with the unified NS EOS proposed by Douchin and Haensel \cite{Douchin01},
which uses the Skyrme SLy4 force.
It can be seen that standard Gogny forces, such as D1M and D1N, predict too low maximum stellar masses, with D1N being unable to  
generate masses above $1.4M_\odot$. We note that this common failure of conventional
Gogny parametrizations \cite{Doan11,Rios14,Gonzalez17} has been cured in the new D1M* force, which, as well as D2 and SLy4, 
is successful in reaching the masses of $2M_\odot$ observed in NSs \cite{Demorest10,Antoniadis13}.
This fact is directly related to the behavior of the EOS in $\beta$-stable matter. As can be seen by
looking at Figs.~\ref{esym}b and \ref{MR}, the stiffer the EOS at high density, the larger the maximum NS mass.
Concerning the new D1M$^*$ force, it predicts a maximum NS mass of $2 M_\odot$ 
with a radius of 10.2 km, and a canonical star of $1.4M_\odot$ with a radius of 11.6 km. 
These values for NS radii are in line with the recent astrophysical extractions of NS sizes
from low-mass X-ray binaries and X-ray bursters that provide radii below 13 km for canonical 
mass stars (see the M-R constraints plotted in Fig.~\ref{MR} \cite{Nattila16,Lattimer14}).

\begin{figure}[b!]{}
\centering
\includegraphics[width=0.85\columnwidth,clip=true]{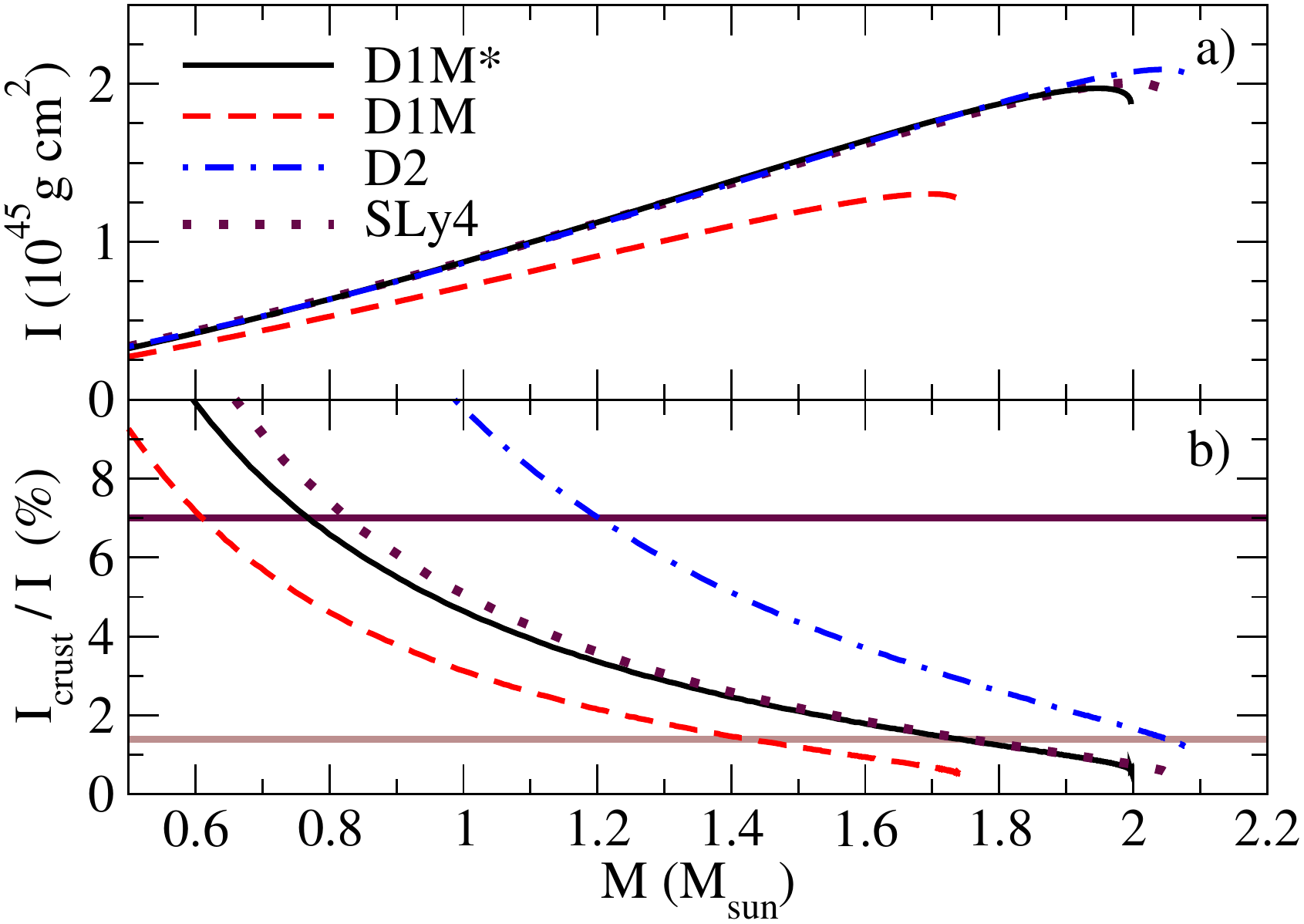}
\caption{
a) Total moment of inertia of the neutron star, 
and b) crustal fraction of the moment of inertia, as functions of the 
total mass of the star. The horizontal lines in (b) depict the bounds 
$I_{\rm crust}/I > 1.4\%$ of \cite{Link1999} and $I_{\rm crust}/I > 7\%$ of 
\cite{Andersson2012} from studying Vela pulsar glitches.}
\label{Icrust}
\end{figure}

We close this section by presenting some results for the moment of inertia of the NS and of its crust.
Astronomical observations of binary pulsars can provide information about the moment of 
inertia of NSs, which in turn may impose constraints on the EOS \cite{Lattimer05}.
In the slow-rotation regime, the NS moment of inertia can be calculated 
by solving simultaneously the TOV equations and the differential equation for the 
moment of inertia in general relativity \cite{Hartle1967}.
Our so-obtained results for the moment of inertia of the star are plotted in Fig.~\ref{Icrust}a.
The predictions of D1M$^*$ and D2 are in close agreement with those from the SLy4 EOS. 
A precise measurement of $I$ is expected within few years for pulsar~A of the double system
PSR J0737-3039 \cite{Lyne:2004cj}.
For the mass $1.34 M_\odot$ of pulsar~A, D1M$^*$, D2 and SLy4 make a prediction of   
$I= 1.29$$\times$$10^{45}$~g\,cm$^{-3}$.

Pulsar glitches 
may be indicative of the fraction of the moment of inertia stored in the crust.
To explain the size of glitches observed in the Vela pulsar, initial studies 
suggested $I_{\rm crust}/I > 1.4\%$ \cite{Link1999}, while $I_{\rm crust}/I > 7\%$ was 
obtained more recently by accounting for entrained neutrons in the crust 
\cite{Andersson2012}. These bounds are shown as horizontal lines in Fig.~\ref{Icrust}b. 
For either bound, the results for $I_{\rm crust}/I$ from D1M$^*$ suggest possible masses 
of the glitching source in between D2 and D1M. The predictions from D1M$^*$ are again 
found to be in consonance with those from the SLy4 EOS.

\section{The fit of the D1M$^*$ interaction and further results}
\label{sec-fit}
To determine the new Gogny interaction D1M$^*$, we have modified the values of the parameters that control  
the stiffness of the symmetry energy while retaining as much as possible the quality of D1M 
for the binding energies and charge radii of nuclei. The basic concept is similar to 
previous literature where families of Skyrme and RMF parametrizations were generated starting from accurate models, 
as for example the SAMi-J \cite{Roca13}, KDE0-J \cite{Agrawal05} or FSU-TAMU \cite{Piekarewicz11,Fattoyev13} families.
In our case, we readjust the eight parameters $W_i$, $B_i$, $H_i$, $M_i$ ($i=1,2$) of the 
finite-range part of the Gogny interaction (\ref{VGogny}),
while keeping the remaining parameters of Eq.\ (\ref{VGogny}) fixed to the values of D1M.
The open parameters are constrained by requiring the same saturation density, energy per particle, compressibility and effective mass 
in symmetric nuclear matter as in the original D1M force, and, in order to have a correct description of 
asymmetric nuclei, the same value of $E_{\rm sym}(0.1)$, i.e., the symmetry energy at density 0.1 fm$^{-3}$.
The last condition is based on the fact that the binding energies of finite nuclei constrain 
the symmetry energy at an average density of nuclei of about 0.1 fm$^{-3}$ more tightly than 
at the saturation density $\rho_0$ \cite{Horowitz:2000xj,Centelles09}.
To preserve the pairing properties in the $S=0$, $T=1$ channel, 
we demand in the new force the same value of D1M for the two combinations of parameters $W_i-B_i-H_i+M_i$ ($i=1,2$). 
Thus, we are able to obtain seven of the eight free parameters of D1M$^*$ as a function of a 
single parameter, which we chose to be $B_1$. This parameter is used to modify the slope $L$ of the symmetry energy at 
saturation and, therefore, the behavior of the neutron matter EOS above saturation, which in turn determines the 
maximum mass of neutron stars. In this way the parameters of the finite-range part of the new interaction
D1M$^*$ are completely determined.
Finally, we perform a small readjustment of the zero-range strength $t_3$ to optimize 
the results for nuclear masses (see Sec.~\ref{sec:Finitenuclei}), which
induces a slight change in the values of the saturation properties of uniform matter.

The parameters of the new force D1M$^*$ are collected in Table \ref{param} and several nuclear matter properties in Table \ref{inm}.
Though the change in the $W_i$, $B_i$, $H_i$, $M_i$ values is relatively large with respect to the D1M values \cite{Goriely09}, 
the saturation properties of symmetric nuclear matter and the symmetry energy at 0.1~fm$^{-3}$  
are basically the same as in D1M (see Table~\ref{inm}).
The mainly modified property is the density dependence of the symmetry energy, with a change in the slope from $L=24.83$ MeV 
to $L=43.18$ MeV, in order to provide a stiffer neutron matter EOS and limiting NS masses of $2 M_\odot$.
The different $L$ value, as we fixed $E_{\rm sym}(0.1)$, implies that 
the symmetry energy $E_{\rm sym}(\rho_0)$ at saturation differs in D1M$^*$ from D1M.
Table~\ref{inm} also reports the nuclear matter properties of the other forces.
It is particularly noticeable that $L$ in D2 ($44.85$ MeV) is fairly larger than 
the values predicted by the D1 family and close to $L$ obtained in D1M$^*$. 

\begin{table}[t!]
\centering
\resizebox{\columnwidth}{!}{%
\begin{tabular}{cccccc}
\hline
D1M$^*$      & $W_i$& $B_i$ & $H_i$ & $M_i$ & $\mu_i$  \\
$i$=1    & $-$17242.0144 & 19604.4056  & $-$20699.9856 & 16408.3344 & 0.50 \\
$i$=2    &    675.3860 &  $-$982.8150  &    905.6650 &  $-$878.0060 & 1.00 \\ \hline
         & $t_3$ & $x_3$ & $\alpha$ & $W_{LS}$&  \\
         & 1561.22           & 1     & 1/3      & 115.36              &  \\ \hline
\end{tabular}
}
\caption{Parameters of the new D1M$^*$ Gogny interaction. 
$W_i$, $B_i$, $H_i$, $M_i$ are in MeV, $\mu_i$ in fm, $t_3$ in MeV\,fm$^4$, $W_{LS}$ in MeV\,fm$^5$, and $x_3$ and $\alpha$ are unitless.}
\label{param}
\end{table}

\begin{table}[t!]
\centering
\resizebox{\columnwidth}{!}{%
\begin{tabular}{lccccccc}
\hline
         &$\rho_0$ & $E_0$ & $K$ & $m^*/m$   & $E_{\rm sym}(\rho_0)$ & $E_{\rm sym}(0.1)$ & $L$  \\ 
         & (fm$^{-3}$) &  (MeV)  &  (MeV) &        & (MeV) &(MeV) & (MeV)\\  \hline
D1M$^*$  &       0.1650          & $-$16.06  & 225.4   &  0.746  & 30.25   & 23.82 & 43.18  \\ 
D1M      &       0.1647          & $-$16.02  & 225.0   &  0.746  & 28.55   & 23.80 & 24.83  \\
D1N      &       0.1612          & $-$15.96  & 225.7   &  0.697  & 29.60   & 23.80 & 33.58 \\
D1S      &       0.1633          & $-$16.01  & 202.9   &  0.747  & 31.13   & 25.93 & 22.43 \\
D2       &       0.1628          & $-$16.00  & 209.3   &  0.738  & 31.13   & 24.32 & 44.85  \\
SLy4     &       0.1596          & $-$15.98  & 229.9   &  0.695  & 32.00   & 25.15 & 45.96 \\\hline
\end{tabular}
}
\caption{Nuclear matter properties predicted by the D1M$^*$, D1M, D1N, D1S and D2 Gogny 
interactions and the SLy4 Skyrme force.}
\label{inm}
\end{table}

\begin{figure}[b!]
\centering
\includegraphics[width=0.85\columnwidth,clip=true]{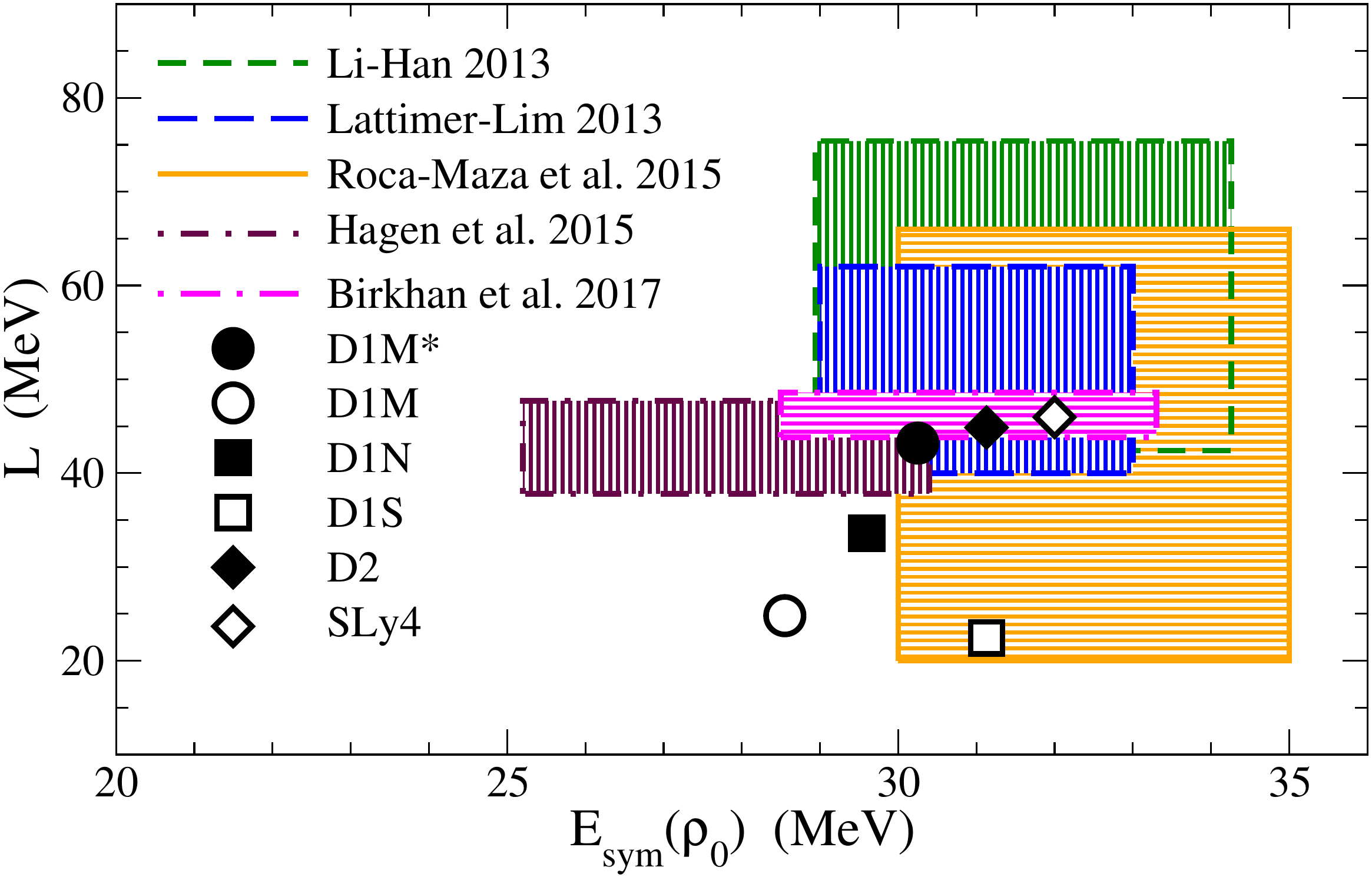}
\caption{Slope $L$ and value $E_{\rm sym}(\rho_0)$ of the symmetry energy at saturation density
for the discussed interactions. The hatched regions are the experimental and theoretical constraints 
derived in \cite{Li:2013ola,Lattimer:2012xj,Roca-Maza:2015eza,Hagen:2015yea,Birkhan:2016qkr}.}
\label{esym_l}
\end{figure}

A few recent bounds on $E_{\rm sym}(\rho_0)$ and $L$ proposed  
from analyzing different laboratory data and astrophysical observations 
\cite{Li:2013ola,Lattimer:2012xj,Roca-Maza:2015eza} and from ab initio nuclear calculations using
chiral interactions \citep{Hagen:2015yea,Birkhan:2016qkr}
are represented in Fig.~\ref{esym_l}. The prediction of D1M$^*$ is seen to overlap with the various constraints.
We note this was not incorporated in the fit of D1M$^*$.
It follows as a consequence of having tuned the density dependence of the symmetry energy of the interaction  
to be able to reproduce heavy NS masses simultaneously with the properties of nuclear matter and nuclei.
D2 and SLy4 also show good agreement with the constraints of Fig.~\ref{esym_l}. 
We observe that the three interactions have an $E_{\rm sym}(\rho_0)$ value of 30--32 MeV
and an $L$ value of about 45 MeV. A similar feature was recently found in the frame of RMF 
models if the radii of canonical NSs are to be no larger than $\sim$13~km 
\cite{Chen:2014mza,Tolos:2016hhl}.~It seems to us a remarkable~fact the convergence of 
mean field models of different nature (Gogny, Skyrme, and RMF) 
to specific values $E_{\rm sym}(\rho_0)$$\sim$30--32 MeV and $L$$\sim$\,45 MeV for the nuclear symmetry energy
when the models successfully describe the properties of nuclear matter and finite nuclei 
and heavy neutron stars with compact stellar radii.

\subsection{Finite nuclei}
\label{sec:Finitenuclei}
One of the goals of the present Gogny D1M$^*$ force is to 
preserve the good performance of D1M in describing nuclear structure features of 
finite nuclei. We have checked that the basic bulk properties of 
D1M$^*$, such as binding energies or charge radii of even-even nuclei, remain 
globally unaltered as compared to D1M. The finite nuclei 
calculations have been carried out with the \mbox{HFBaxial} code
\cite{HFBaxial} using an approximate second-order gradient method to 
solve the HFB equations \cite{rob11} in a harmonic oscillator (HO) 
basis. The code preserves axial symmetry but is allowed to break 
reflection symmetry. It has already been used in large-scale 
calculations of nuclear properties with the D1M force, as e.g.\ in Ref.\ \cite{Rob11b}.
We carried out HFB calculations for 620 even-even nuclei of the 2012 AME \cite{Audi12} using the \mbox{HFBaxial} code.
First, the potential energy surface (PES) as a function of the quadrupole moment $Q_{20}$ 
is computed to select the lowest-energy minimum, which is subsequently
used to start an unconstrained calculation to obtain the true HFB ground state.
The binding energy is obtained by subtracting to the HFB energy the rotational energy correction,
as given in Ref.\ \cite{RRG00}. The ground-state calculation is repeated with an
enlarged basis containing two more HO major shells and an extrapolation
scheme to an infinite HO basis is used to obtain the final binding energy \cite{Hilaire.07,Baldo13}.
In our framework, the zero-point energy (ZPE)
of quadrupole motion used in the original fitting of D1M \cite{Goriely09} is not taken into account
because it requires considering $\beta$-$\gamma$ PES and solving
the five-dimensional collective Hamiltonian for all the nuclei. This is still an enormous task
and we follow a different strategy where the quadrupole ZPE is replaced by a constant binding energy shift. This is 
somehow justified as in our experience the ZPE shows a weak mass dependence
(see \cite{Rob15} for an example with the octupole degree of freedom). The energy shift is fixed
by minimizing the global rms deviation, $\sigma_{E}$, for the 
known binding energies of 620 even-even nuclei \cite{Audi12}.
With a shift of 2.7 MeV we obtain for D1M a $\sigma_{E}$ of 1.36 MeV,
which is larger than the 798 keV reported for D1M in \cite{Goriely09} including also odd-even and odd-odd nuclei. 
The result is still satisfying and gives us confidence on the procedure followed.
Using the same approach for D1M$^{*}$ we obtain a $\sigma_{E}$ of
1.34 MeV (with a shift of 1.1 MeV), which compares favorably with our $\sigma_{E}$ of 1.36 MeV for D1M.

\begin{figure}[t!]{}
\centering
\includegraphics[width=0.85\columnwidth]{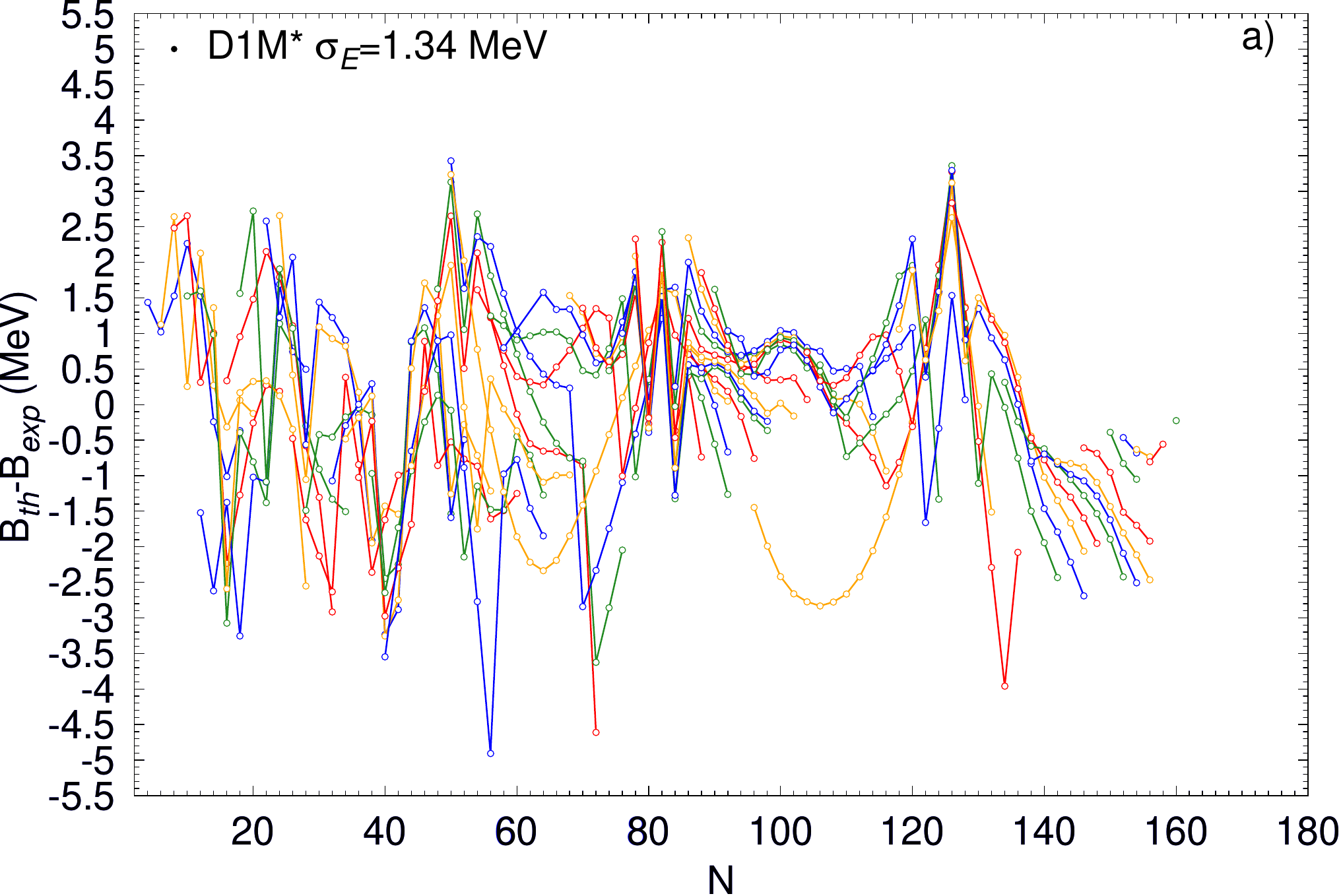}
\includegraphics[width=0.85\columnwidth]{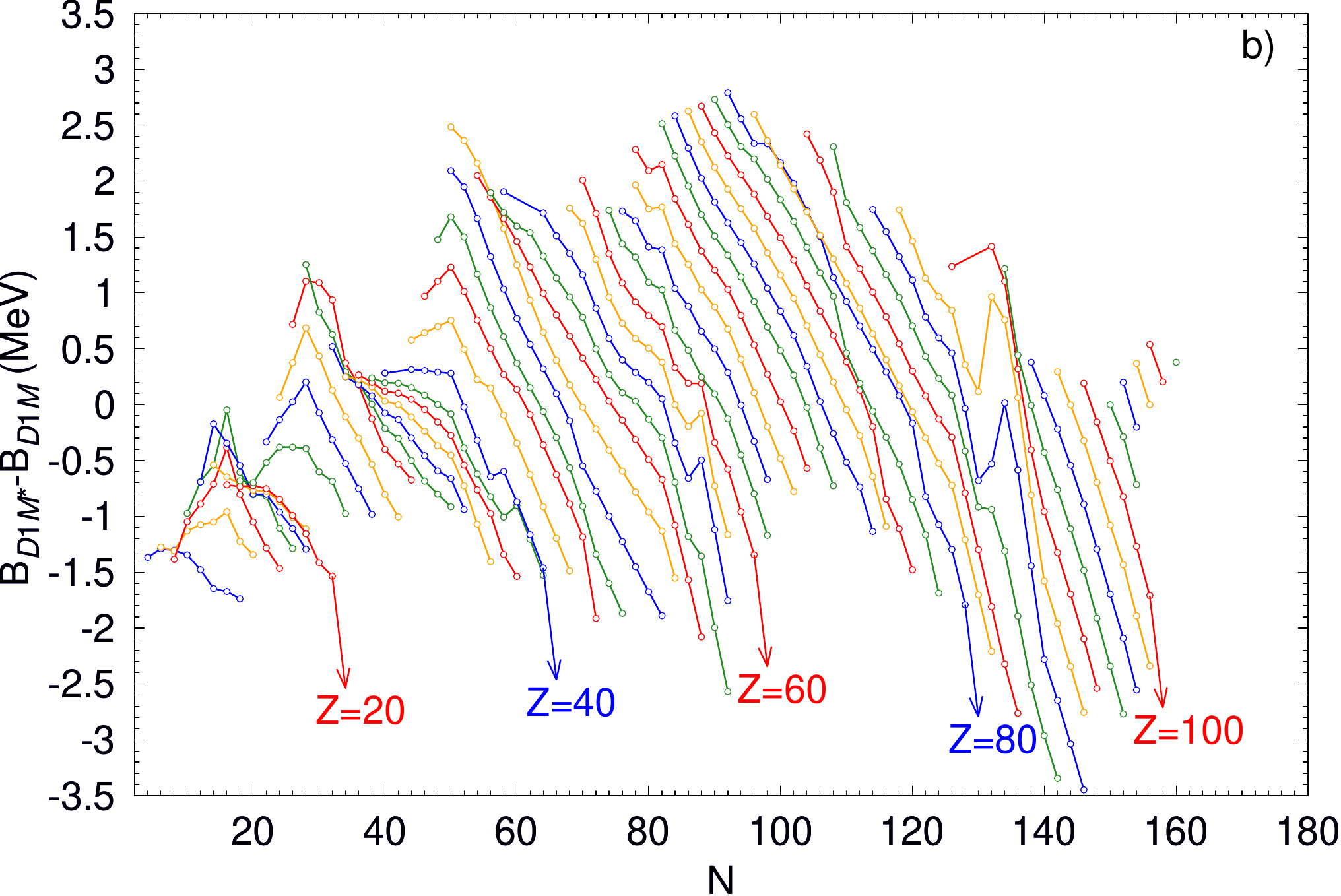}
\caption{a) Binding energy differences in 620 even-even nuclei between computed and 
experimental values \cite{Audi12}
for the new D1M$^*$ force, as a function of neutron number $N$.
b) Binding energy differences between the theoretical predictions of D1M$^{*}$ and D1M
for 818 even-even nuclei.\label{Bfn}}
\end{figure}

The differences $\Delta B=B_\textrm{th}-B_\textrm{exp}$ between the binding energies of
D1M$^*$ and the experimental values for 620 even-even nuclei belonging to different isotopic 
chains are displayed in Fig.~\ref{Bfn}a against the neutron number $N$. 
The $\Delta B$ values are scattered around zero and show no drift with increasing $N$.
The agreement between theory and experiment is specially good for medium-mass and heavy
nuclei away from magic numbers and deteriorates for light nuclei, 
as may be seen from the partial $\sigma_{E}$ deviations given in Table~\ref{prms}.
From the partial $\sigma_{E}$ values of Table~\ref{prms}, we also conclude that the
closeness in the total $\sigma_{E}$ of D1M and D1M$^{*}$ involves subtle
cancellations that take place all over the nuclear chart.
We plot the differences in binding energy predictions between D1M$^*$ and D1M in Fig.~\ref{Bfn}b 
against $N$ for 818 even-even nuclei.
\begin{table}[t!]
\centering
\resizebox{0.6\columnwidth}{!}{%
\begin{tabular}{lccc}
\hline
           & $A \leq 80$    &  $80 < A \leq 160$    & $A > 160$ \\ \hline
D1M$^{*}$  & 1.55        &  1.31                &   1.26    \\
D1M        & 1.82        &  1.12                &   1.29    \\ \hline
\end{tabular}
}
\caption{Partial rms deviation (in MeV) 
from the experimental binding energies \cite{Audi12} in even-even nuclei, computed in the given mass-number intervals.}
\label{prms}
\end{table}
A clear shift is observed as $N$ increases within an isotopic chain.
It is a direct consequence of the different density dependence of the symmetry energy in the two interactions. A similar 
behavior can be observed in a recent comparison \cite{Pillet17} between D2 and D1S. It is also interesting to note that
the results for neutron radii show a similar isotopic drift as the
 binding energies. Namely, the difference $r_\mathrm{D1M^{*}}-r_\mathrm{D1M}$
(where $r$ is the rms radius computed from the HFB
wave function) increases linearly with $N$ for the neutron radii, 
whereas it remains essentially constant with $N$ for the proton radii. This is again
a consequence of the larger slope $L$ of the symmetry energy in D1M$^{*}$ \cite{Brown00,Centelles09}. 
All these effects, as well as quadrupole, octupole and fission deformation properties
of the new Gogny D1M$^{*}$ force will be analyzed in detail in forthcoming work.
There are already indications of a good performance of D1M$^{*}$ in describing the basic 
parameters of fission.

\section{Summary and conclusions}
The existence of neutron stars with large masses of $2M_\odot$ 
\cite{Demorest10,Antoniadis13} has been exploited in recent years to select equations 
of state satisfying astrophysical evidence. 
The most successful forces of the D1 family of the finite-range Gogny interaction, 
which plays an important role in nuclear physics, have soft symmetry energies and fail 
to produce heavy enough stellar masses.
We propose a way to reparametrize a Gogny force where, while preserving 
the description of nuclei, the slope of 
the symmetry energy can be modified as to make the EOS of $\beta$-stable matter stiffer to 
obtain NS masses of $2M_\odot$.
The D1M force \cite{Goriely09} is susceptible to being used in this procedure,
but not D1S and D1N that are too far from the $2M_\odot$ target. 
We find that the new set of parameters, denoted as D1M$^*$, is reconciled
with the prediction of $2M_\odot$ stars and performs at the same level 
as D1M in all aspects of finite nuclei analyzed in this work. Stellar properties from 
D1M$^*$, such as the \mbox{M-R} relation and the moment of inertia,
are in good agreement with the results from the Douchin-Haensel SLy EOS \cite{Douchin01}.
Although much more work is required to assess the performance of D1M$^{*}$, as e.g.\ in 
fission studies, we conclude that it represents a promising alternative in the 
description of nuclei and at the same time has the right properties to study exotic 
astrophysics scenarios such as NSs.
In order to complete a unified NS EOS in the realm of a finite-range interaction, 
which allows for a description of the pairing channel free of divergences,
it will be worth computing the structure of the NS crust with the new D1M* Gogny force.

\section*{Acknowledgments}
The work of LMR was supported by Spanish Ministry of Economy and Competitiveness (MINECO) 
Grants No FPA2015-65929-P and FIS2015-63770-P. 
C.G., M.C., and X.V. were partially supported by Grant FIS2014-54672-P from MINECO and FEDER,
Grant 2014SGR-401 from Generalitat de Catalunya,
and Project MDM-2014-0369 of ICCUB (Unidad de Excelencia Mar\'{\i}a de Maeztu) from MINECO.
C.G. also acknowledges Grant BES-2015-074210 from MINECO.

\section*{References}
\bibliography{mybibfile.bib}

\end{document}